\documentclass{aa}
\usepackage{psfig}
\def\ergs{\rm \ erg \, s^{-1}}

\def\cmdue {\rm \ cm^{-2}}

\def\cts {\rm \ count \, s^{-1}}

\begin{document}

\title{The Brera Multi-scale Wavelet ROSAT HRI source catalogue
(BMW-HRI)}

\author{Maria Rosa Panzera\inst{1}
 \and Sergio Campana\inst{1}
 \and Stefano Covino\inst{1}
 \and Davide Lazzati\inst{1,2}
 \and Roberto P. Mignani\inst{3}
 \and Alberto Moretti\inst{1}
 \and Gianpiero Tagliaferri\inst{1}
} 

\institute{INAF -- Osservatorio Astronomico di Brera, via E. Bianchi 46, I--23807
Merate (LC), Italy
\and 
Institute of Astronomy, University of Cambridge, Madingley Road,
CB3 0HA Cambridge, England
\and
European Southern Observatory, Karl Schwarzschild Strasse 2, D85740 Garching bei M\"unchen, Germany}

\date{Received date / accepted date}
\offprints{M.R.~Panzera: panzera@mera\-te.mi.astro.it}

\titlerunning{The BMW-HRI source catalogue}
\authorrunning{Panzera et al.\ }
                          
\abstract
{
We present the Brera Multi-scale Wavelet ROSAT HRI source catalogue
(BMW-HRI) derived from all ROSAT HRI pointed observations with 
exposure time longer than 100 s available in the 
ROSAT public archives.
The data were analyzed automatically using a wavelet 
detection algorithm suited to the detection and characterization of both 
point-like and extended sources.
This algorithm is able to detect and disentangle sources in very crowded fields
and/or in presence of extended or bright sources.
Images have been also visually inspected after the analysis
to ensure verification.
The final catalogue, derived from 4,303 observations, consists of 29,089 sources detected 
with a detection probability of $\geq 4.2 \sigma$. 
For each source, the primary catalogue  entries provide  name, position,
count rate, flux and extension along with the relative errors.
In addition, results  of cross-correlations with  existing
catalogues  at different  wavelengths (FIRST, IRAS, 2MASS and GSC2)
are also reported.
All these information are available on the web via the DIANA Interface.
As an external check, we compared our catalogue with the
previously available ROSHRICAT catalogue (both in its short and long
versions) and we were able to recover, for the short version, $\sim\,90\%$ of the entries.
We computed the sky coverage of the entire HRI data set by means of simulations.
The complete BMW-HRI catalogue provides a sky coverage of 732 deg$^{2}$ down to a 
limiting flux of $\sim 10^{-12} \ergs \cmdue$ 
and of 10 deg$^{2}$ down to  $\sim 10^{-14} \ergs \cmdue$.
We were able to compute the cosmological log(N)-log(S) distribution
down to a flux of $\simeq 1.2 \times 10^{-14} \ergs \cmdue$.
}
\maketitle

\keywords{catalogues -- X--rays: general}

\section{Introduction}

Since the early phases of X--ray astronomy, source catalogues have
been regularly compiled either from systematic sky surveys or from the
collection  of  serendipitous  observations, especially  when  imaging
X--ray telescopes became available (e.g. {\it Einstein} 
EMSS, Gioia et al. \cite{gioia90}; EXOSAT HGSC, Giommi et al. \cite{giommi91}; 
ROSAT WGA, White, Giommi \& Angelini \cite{white94}; 
ROSAT SRC, Zimmermann \cite{zimmermann94}; ASCA SIS, Gotthelf \& White
\cite{gotthelf97}). 
Depending on the telescope field of view and mission lifetime, these
surveys covered $1-10\%$ of the sky and allowed to pursue the statistical studies
on virtually all classes of X--ray emitting sources.

Observations with  the ROSAT PSPC  produced a  number of
catalogues; besides the RASS-BSC  (Voges et al. \cite{voges99}), derived  from the
ROSAT All Sky Survey, there are the WGA (White, Giommi, Angelini \cite{white94}) and the 
ROSPSPC (ROSAT Team 2001), derived  from  the collection of all PSPC pointed 
observation. 
On the contrary the large database of ROSAT High Resolution Imager (HRI)
observations has been only marginally exploited.
Rather recently (August 2001), a full catalogue based on the Standard Analysis
Software System (SASS) has been released (ROSHRICAT, ROSAT Team 2001). 
This catalogue contains arcsecond positions and count rates for automatically 
and visually inspected sources, including 13,452 high confidence (S/N $>4$) 
detections, from 5,393 public ROSAT HRI observations
covering $1.94\%$ of the sky.
However, the SASS suffers from some limitations (e.g. 331 ``obvious'' sources
were added manually since they were not detected by the SASS), not least the
fact that only bright sources can be securely detected.\par

The HRI on board the ROSAT satellite is a microchannel
plate detector with an octagon--like shape field of view 
(with $\sim 19'$ radius) that reveals single X--ray photons providing information
on their positions and arrival times.
The HRI Point Spread Function (PSF) as measured on--axis is of about 5 arcsec FWHM, i.e. a factor
of $\sim 4$ better than the one of ROSAT-PSPC (and just a factor of $\sim 5$ worse than Chandra).
The sharp core of the HRI PSF allows to detect and disentangle 
sources in very crowded fields and to detect extended emission on a small 
angular size.
On the other hand, the HRI has a very crude spectral resolution in the
0.1--2.4 keV energy band (for more details see Prestwich et al. \cite{prestwich96}), it
is less efficient than the PSPC (a factor of 3 to 8 for a plausible range
of incident spectra) and it has a higher instrumental background.
The ROSAT satellite, its X--ray telescope, and the HRI detector have been
described in detail by Pfeffermann et al. (\cite{pfeffermann86}),
Zombeck et al. (\cite{zombeck90}), Tr\"umper et al. (\cite{trumper91}) and 
David et al. (\cite{david98}).

For these reasons, we decided to reanalyze the entire HRI data set
with a dedicated source detection algorithm based on the wavelet
transform (Lazzati et al. \cite{lazzati99}; Campana et al. \cite{campana99}). 
The outcome is a new source 
catalogue named Brera Multi-scale Wavelet HRI (BMW-HRI). 
Here we present the results of our analysis. In section 2 we resume the main 
characteristics of the detection algorithm. In section 3 
we discuss the selection of the ROSAT HRI fields used and we describe the catalogue. 
In section 4 we present the cosmological Log(N)-Log(S) distribution computed from
our catalogue. In section 5
we made a comparison with the ROSHRICAT catalogues. In section 6
we describe the cross-correlations between the BMW-HRI and other catalogues at various 
wavelengths. Conclusions and catalogue accessibility are reported in section 7.

\section{Wavelet detection algorithm}

The analysis and source detection of HRI images, together with 
the simulations carried out to test the detection pipeline, are 
extensively described in Campana et al. (\cite{campana99}) and Lazzati et al. 
(\cite{lazzati99}).
In this case we decided to run the detection algorithm 
with a single significance threshold 
for the sources ($\sim 4.2\,\sigma$) corresponding to a contamination
of 0.4 spurious sources per field.
We remark here that one of the most interesting features of the wavelet analysis
is the possibility of characterizing the source extension
(see sub-section 3.2 and Campana et al. 1999).
The data were retrieved from the MPE and GSFC public ROSAT databases.
Our data processing pipeline analyzes the FITS event files, as produced through
the SASS procedure, 
and the ancillary files that include orbit and spacecraft pointing and 
other engineering or housekeeping information.
From the analysis of each observation
we derived a catalogue of sources with
position, count rate, extension, along with the relative errors, as well as 
ancillary information about the observation itself and source fitting.
The count-to-flux conversion factor was determined assuming as reference a Crab spectrum
(power law with photon index 2.0). A conversion factor was computed both for 
a low column density ($5\,\times\,10^{19}\cmdue$) 
and for the full galactic value.
We also applied the corrections due to the 
vignetting, PSF modelling (we considered a Gaussian fit), plate scale (the nominal 
pixel size was $0.5''$ reduced to $0.4986''$ after detailed observations on 
the Lockman hole field, see Hasinger et al. \cite{hasinger98}) and PSF asymmetry at large 
off--axis angles (see Campana et al. \cite{campana99}).

\begin{figure}
\psfig{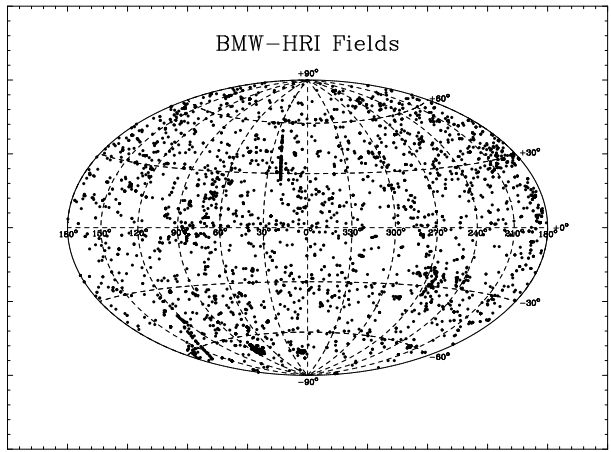}
\caption{Aitoff projection in galactic coordinates for 
the whole set of fields used to construct the BMW-HRI catalogue (4,303 observations).}
\end{figure}

\begin{figure}
\psfig{figure=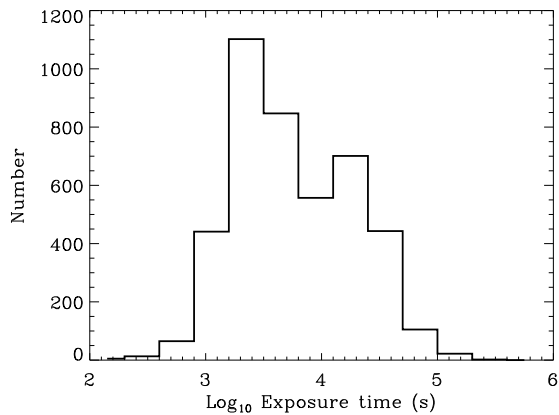,width=8cm}
\caption{Distribution of the exposure times for the 4,303 ROSAT-HRI 
observations considered.}
\end{figure}

\section{The BMW-HRI catalogue}

\subsection{Selection of the ROSAT HRI fields}

The starting point for the BMW-HRI catalogue production was the sample of
4,507 ROSAT HRI fields published till December 2001
with exposure time longer than 100 s.
We did not include in our selection the 341 fields pointed on supernova 
remnants (fields with identification number, ROR, starting with 5) due to 
the large number of bright overlapping spots in which the remnant is splitted. 
This despite the fact that the central source of the (e.g) Cas A remnant 
is clearly detected as point source (Pavlov et al. \cite{pavlov00}). We also did not 
consider the 598 calibration observations 
(fields with ROR starting with 1). 
All the 4,507 fields were analyzed self-consistently with the detection algorithm 
described in section 2.
Of the 4,507 fields analyzed, 204 ($\sim 5\%$) were discarded due to problems 
in the construction of the exposure map, or in the reconstruction of the aspect 
attitude, or due to problems during the analysis.
In particular, several fields affected by the spacecraft wobbling
during the exposures were discarded because bright sources appeared
artificially elongated and therefore splitted into at least two 
sources by the detection algorithm. 
Moreover, very confused fields  like, e.g., the ones targetted to
Eta Carinae, were discarded.
Finally, a few fields pointed on planets and comets were discarded due to their proper 
motion which induces multiple detections across the images. 
Therefore, we ended up with 4,303 fields.
As our detection threshold corresponds to 0.4 spurious sources per fields,
we expect $\sim\,1,721$ spurious sources over all the fields.
In 146 fields no sources were detected, 
due to the too short exposure time. 
Fig. 1 shows the Aitoff projection in galactic coordinates for the 
4,303 observations.

\begin{figure}
\psfig{figure=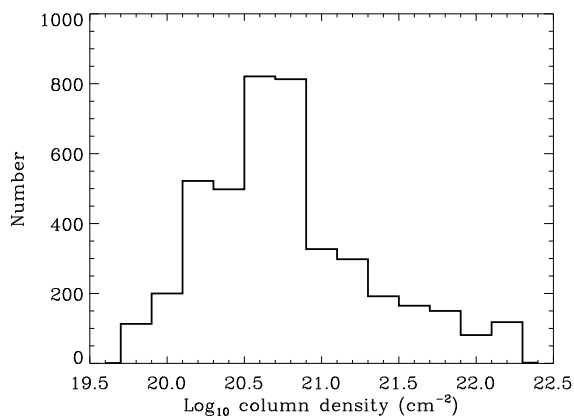,width=8cm}
\caption{Distribution of the galactic column density after Dickey \& Lockman 
(\cite{dickey90}) for the 4,303 ROSAT-HRI observations considered.}
\end{figure}

\begin{figure}
\psfig{figure=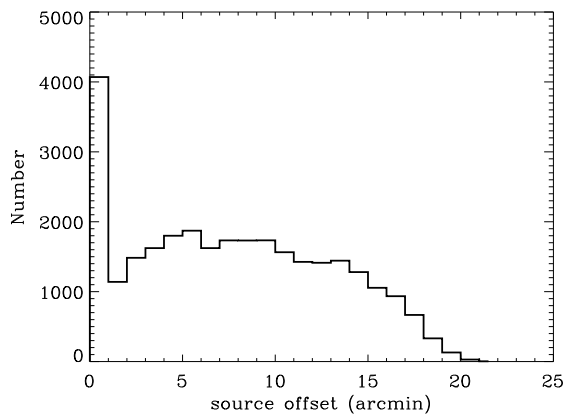,width=8cm}
\caption{Distribution of the off--axis angles for the 29,089 BMW-HRI sources.
The plot shows the typical trend due to: targets plus increasing of the collecting 
area together with the decreasing of the sensitivity with the off-axis.
}
\end{figure}

The distribution of the exposure times can be roughly described by a double Gaussian 
distribution with $\sim 3$ and $\sim 25$ ks as centroid peaks (see Fig. 2). 
In Fig. 3 we plot the distribution of the galactic hydrogen column density after
Dickey \& Lockman (\cite{dickey90}).

\subsection{Catalogue characteristics}

The BMW-HRI catalogue consists of 29,089 sources down to a limiting
ROSAT HRI count-rate of  $\sim 10^{-4} \cts$ covering $\sim 1.8\%$ of the 
sky.
A detailed description of all the BMW-HRI parameters
together with the ones derived from the cross-correlations (see section 6)
can be found in the Appendix A (table A.1, A.2, A.3, A.4 and A.5).
The catalogue lists all the sources detected in all the 4,303
fields analyzed, i.e. we did not associate sources
detected in different observations of the same area of the sky.
An estimate of the number of independent sources in the catalogue can be 
obtained compressing the number of sources using an error on the source position of
$10''$ (a conservative limit dictated by the uncertainties connected with the 
boresight correction). We found 20,453 independent objects.

In Fig. 4 we plot the distribution of source off--axis angles showing the 
typical trend due to the increase of the collecting area
together with the decreasing of the sensitivity with off--axis.
Clearly, the peak at zero off--axis is due to pointed sources.

One of the most important characteristics of wavelet algorithms is the 
ability to determine the source extension, i.e. the scale of the wavelet transform
where the S/N is maximized after the application of the fitting refinement procedure
(see Lazzati et al. \cite{lazzati99}), 
and, if a criterion is given, also to disentangle point and extended sources
(see Campana et al. \cite{campana99}). 
To assess the source extension criterion 
we considered all sources detected in the observations that have a star(s) as a 
target (ROR number beginning with 2) and that were available in the public 
archives in a preliminary phase of our catalogue: 6,013 sources in 756 HRI fields.
The distribution of the source extension as a function of the 
source off--axis angle has been divided into bins of 1 arcmin each.
To each bin we then applied a $\sigma-$clipping algorithm
to discard iteratively truly extended sources and to derive the mean value of the
source extension in the bin for pointed sources.
We then determined the $3\,\sigma$ dispersion on the mean for each bin.
The mean value plus the $3\,\sigma$ dispersion provides the threshold for the 
source extension. 
We conservatively classify a source as extended if it lies more than $2\,\sigma$ 
from this limit (i.e. if the source extension error bar lies more than twice 
from the $3\,\sigma$ limit described above).
Combining this threshold with the $3\,\sigma$ on the intrinsic dispersion,
we obtain a $\sim\,4.5\,\sigma$ confidence level for the extension classification
(see also Rosati et al. \cite{rosati95}).
In Fig. 5 we plot the distribution of the source extension versus off--axis angle 
for the 29,089 BMW-HRI sources (small dots). The solid line in Fig. 5 represents
the mean value of source extension for pointed sources as described above, while 
the dashed line is the $3\,\sigma$ dispersion on this mean.
Open squares in Fig. 5 represent sources we classified as extended, i.e. with
a confidence level for the extension classification of $\sim\,4.5\,\sigma$. 

We end up with 2,717 extended sources (open squares in Fig. 5) containing supernova 
remnants, galaxies, cluster of galaxies etc. (as well as blending of nearby sources). 
The distributions of source extensions for the truly extended sources, the full sample
and only the high-galactic latitude ($|b| > 20^{\degr}$) one (2,139 sources), are 
reported in Fig. 6.
We assumed that all the high-galactic latitude
extended sources are extragalactic in nature.
In the plot we also report the on--axis angular resolution of the ROSAT PSPC.
X--ray extensions are calculated subtracting in quadrature the relative 
PSF extension at a given off--axis angle (solid line in Fig. 5). 
The extended sources were used to select a list of candidate X--ray selected 
cluster of galaxies that we then studied with optical follow-up 
(Moretti et al. \cite{moretti02} in preparation; Guzzo et al. \cite{guzzo02} 
in preparation).  

\begin{figure}
\psfig{figure=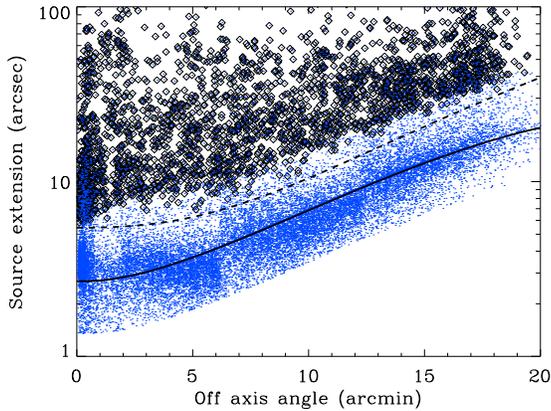,width=8cm}
\caption{Distribution of BMW-HRI source extensions. The solid line 
represents the HRI PSF, whereas the dashed line the $3\,\sigma$
limit for point sources as described in the text. Small dots represent the 
extension for all sources. Open squares 
denote truly ($\sim 4.5\,\sigma$) extended sources i.e. sources that lies more
than $2\,\sigma$ from the dashed line (2,717 sources).}
\end{figure}

\begin{figure}
\psfig{figure=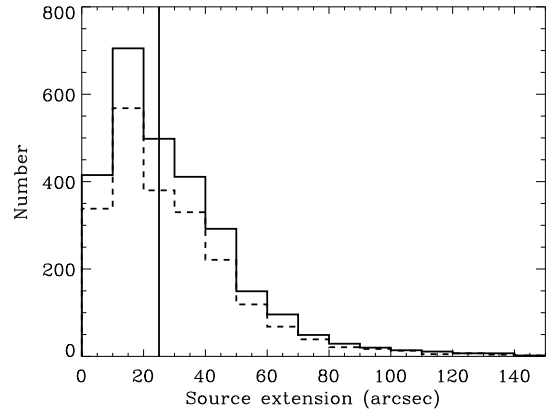,width=8cm}
\caption{Distribution of BMW-HRI source extensions for the 2,717 
extended sources (solid line). 
The dashed line represents the distribution of 
the high-galactic latitude ($|b| > 20^{\degr}$) extended sources (2,139 objects), 
whereas the vertical solid line
the on--axis angular resolution of the ROSAT PSPC.
The X--ray extensions are calculated subtracting in quadrature
the intrinsic PSF width.
}
\end{figure}

As the sensitivity of the HRI instrument is not uniform over the entire field of view,
for a given limiting flux the surveyed area does not coincide with the
detector one but it is generally smaller. We calculated the
sky coverage of the entire survey as a function of the flux 
(calculated with the full column density) by means of simulations.
To this aim we performed extensive Montecarlo tests using the detection
procedures on the simulated data for 12 observations
with different exposure times (i.e. 1,680 s, 2,656 s, 4,976 s, 10,000 s, 16,080 s,
20,664 s, 27,488 s, 42,672 s, 52,880 s, 80,288 s, 107,712 s and 200,272 s).
For each field of the survey we derived a completeness
function that gives the detection probability as a function of flux
and position of point-like sources. Because of the worsening of the PSF, 
within the same field, the detection probability of a given flux
decreases for increasing off--axis angle.
Therefore, for each observation the sky coverage is the
integral of the completeness function over the field of view.
Moreover, each observation has different completeness
function depending on the exposure time and on the column density
values. The sky coverage of the whole survey is the sum
of the contributions of each single field.
Some of the fields used to build the BMW-HRI catalogue cover the same
area of the sky: in this case in the calculation of the total sky coverage
we considered only the contribution from the longest observation
performed over that area.
All the procedures of data simulations and the features of the Montecarlo
tests are fully described in Moretti et al. (\cite{moretti02} in preparation).
The complete sky coverage for point-like sources is shown in Fig. 7
(for extended sources see Moretti et al. \cite{moretti02} in preparation).
The maximum area of the survey is $\sim 732$ deg$^{2}$ and corresponds to
fluxes above $\sim\,10^{-12} \ergs \cmdue$. At  $10^{-13} \ergs \cmdue$ the surveyed 
area is $\sim 314$ deg$^{2}$  and $\sim 10$ deg$^{2}$ at  $10^{-14} \ergs \cmdue$.

\begin{figure}
\psfig{figure=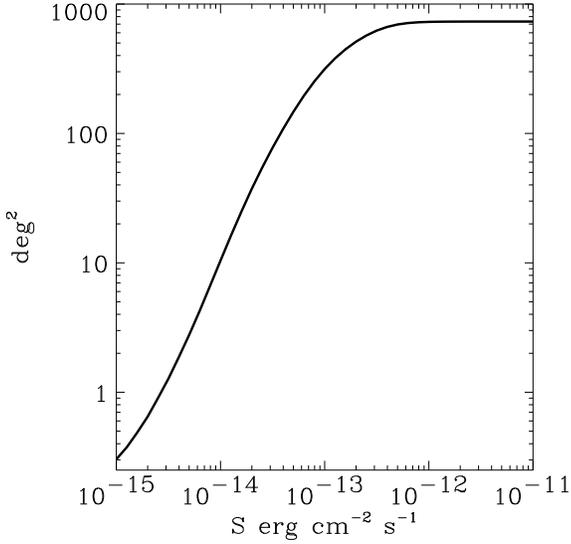,width=8cm}
\caption{Sky coverage for point-like sources of the entire survey as a function of the flux
(full column density): the maximum area of the survey is $\sim 732$ deg$^{2}$ and 
corresponds fluxes above $10^{-12} \ergs \cmdue$. At  $10^{-13} \ergs \cmdue$ the 
surveyed area is $\sim 314$ deg$^{2}$ and $\sim 10$ deg$^{2}$ at  
$10^{-14} \ergs \cmdue$.}
\end{figure}

\begin{figure}
\psfig{figure=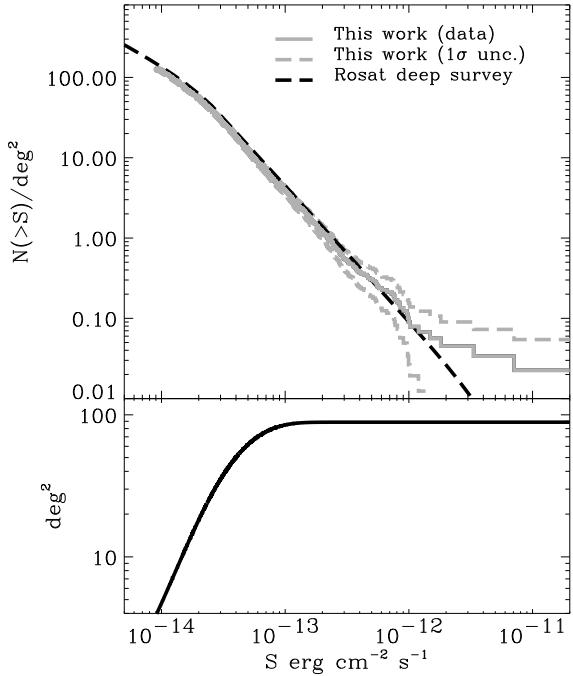,width=8cm}
\caption{Upper panel: comparison between the BMW-HRI log(N)-log(S)
derived for the cosmological sources (bold gray solid line) and the ROSAT Deep 
Survey distribution (dashed line; Hasinger et al. \cite{hasinger98}).
The two thin gray solid lines represents our 1 $\sigma$ uncertainties. 
The comparison is made in the bright part of ROSAT Deep Survey (see text); 
lower panel:
sky coverage of the 501 selected fields as a function of the flux (see text).}
\end{figure}

\section{Log(N)-Log(S)}

In order to compute the integral flux distribution (log(N)-log(S))
of cosmological sources we used a subsample of the catalogue.
Firstly, we selected all high galactic latitude fields ($|b| > 30^{\degr}$)
observed for more than 5 ks.
Then, we filtered out those containing extended
sources (like NGC objects and galaxies clusters),
Magellanic Clouds targets or crowded areas (e.g. M31). Finally, from the
remaining fields, we choose sources with off--axis angles between 3 and 15 arcmin. 
To summarize, we restricted to 501 fields ($\sim 90$ deg$^{2}$) and 3,161 sources.
The survey is inhomogeneous (because of the worsening of the PSF with the off--axis) 
thus in the computation of the flux
distribution different sources have different weights.
The weight is defined as the inverse of the area in which the source has a
non-zero probability of being detected.
Our results are compared in Fig. 8 with those of the
ROSAT Deep Survey (Hasinger et al. \cite{hasinger98}).
We derive our distribution down to $\simeq 1.2 \times 10^{-14} \ergs \cmdue$ where the 
surveyed area corresponds to $8 \%$ of the total ($\simeq 8$ deg$^{2}$).
The ROSAT Deep Survey extends to fainter fluxes and it is well fitted
by a broken power law, with the break at $\sim 2 \times 10^{-14} \ergs \cmdue$
(Hasinger et al. \cite{hasinger98}).
The BMW-HRI distribution is very similar both in steepness and in normalization
to the ROSAT Deep Survey, but extends to brighter fluxes, a factor of 2
after the break point.
In order to compare the two distributions, we exclude from the BMW-HRI distribution
the fainter fluxes ($< 7 \times 10^{-14} \ergs \cmdue$; this is why we cannot constrain
the power law below the break with such few points) and we compared the BMW-HRI 
distribution with the bright part of the ROSAT Deep 
Survey.
By means of a maximum likelihood minimization fit we find that, assuming a single 
power law for the differential distribution, the best fit for the exponential
is given by $\alpha = -2.75\,\pm\,0.11$ with a normalization of
$229.8^{+69}_{-60}$ (in units of $10^{-14}$).
This value is in very good agreement with the bright part of the
ROSAT Deep Survey flux distribution ($\alpha = -2.72$ and a
normalization of 238.1; Hasinger et al. \cite{hasinger98}). 

\section{Comparison with the ROSHRICAT catalogues}

We compared the BMW-HRI with 
the ROSAT source catalogue of pointed observations with the High Resolution
Imager (ROSHRICAT/1RXH, ROSAT Team 2001).
This catalogue, derived by reprocessing the public HRI
dataset (a total of 5,393 pointings covering 1.94$\%$ of the sky) through the SASS, 
provides arcsecond positions and count rates for 131,902 sources.  
This version includes detections which were classified as false after a visual inspection
(``f''  detections), multiple detections of the same source within the same 
observation (``u'' detections) and 331 obvious sources which were not detected by the 
SASS and added manually (see Appendix A.2).  After removing
``u'' and ``f'' detections, 56,401 entries are left
(ROSHRICAT long version).
Additionally, applying a $S/N\,>4$ yields to 13,452 high 
confidence detections (ROSHRICAT short version).\par

We compared the BMW-HRI catalogue with the ROSHRICAT, both in its
short and its long version (hereafter ROSHRICAT-short and
ROSHRICAT-long, respectively) by cross-correlating the entries in the
two catalogues.   
The results of these cross-correlations are
summarized in this section while a detailed description of specific
checks is given in Appendix A.
For consistency, we applied to
the ROSHRICAT catalogues the same selection criteria applied in the
compilations of the BMW-HRI (see section 3). Moreover, we filtered out
from the ROSATHRICAT catalogues the 331 entries which were not
detected by the SASS (no flux information),
plus some detections with a wrong declination (4 sources for the short 
version and 64 sources for the long version).
We remain with 10,708 and 43,252 entries for the ROSHRICAT-short
and long respectively.
To compute the cross-correlation radius we first calculated the positional error
corresponding to the 95$\%$ of the sources both for the BMW-HRI and for the ROSHRICAT
catalogues.
By adding in quadrature the two errors we obtain radii of 8 and 12
arcsec, to be used for the cross-correlations with the ROSATHRICAT
short and long, respectively.
As the number of the ROSHRICAT sources that cross-correlate with each
BMW-HRI sources can be, in several cases, more than one, we decided to
choose the nearest ROSHRICAT source (minimum distance approach).
\par

\subsection{Comparison with the ROSHRICAT short version}

We  first  cross-correlated  the  BMW-HRI  (29,089  entries)  with  the
ROSATHRICAT-short (10,708 entries) using  a cross-correlation radius of
8 arcsec, as determined above, and finding 12,442 associations.
This means that a single ROSHRICAT-short source matches more than one BMW-HRI object.
We remember that, for our catalogue, we did not associate objects
detected in different observations of the same area of the sky.
The number of ROSHRICAT single sources in the cross-correlation is 9,670.
By shifting the coordinates of 3  arcmin we found that the probability of 
mismatch is of the order of $\sim$ 0.6$\%$ (that is 73 mismatches). 
Comparing the ROSHRICAT-short count rates with the BMW-HRI ones
we found that a fraction of 87$\%$ has count rates equal within a factor of 2.
1,038 ROSATHRICAT-short sources are found without a correspondence 
in the BMW-HRI source list. 
In order to check only firmly detected sources we considered
objects with S/N $\ge\,5$ (378 entries).
We found that:
$\sim\,62\%$ have a BMW-HRI association within 30 arcsec;
$\sim\,32\%$ are spurious sources while 
$\sim\,6\%$ are of ambiguous interpretation and in some cases could be true sources 
lost by our algorithm (a more extensive discussion is given in Appendix A.1).

The number of BMW-HRI sources without a counterpart in the ROSHRICAT short version
is 16,647. We discuss the sources lost in the ROSHRICAT catalogues
in section 5.2 since a source lost in the short version can have
a counterpart in the long version.

In Fig. 9 we show the distributions of the angular separation ($r$ in arcsec) between 
ROSHRICAT counterpart position and BMW-HRI position for 
the 12,442 cross-correlated objects.

\begin{figure}
\psfig{figure=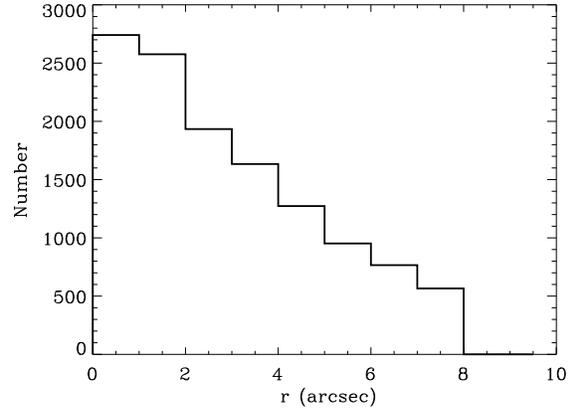,width=8cm}
\caption{ROSHRICAT - BMW-HRI: distribution of the angular separation ($r$) between 
ROSHRICAT-short (see text) and BMW-HRI positions for the 12,442  
cross-correlated objects.}
\end{figure}

The BMW-HRI source count-rates for the common entries are plotted 
(in logarithmic scale) in Fig. 10 vs. the corresponding values from the 
ROSHRICAT-short.

\begin{figure}
\psfig{figure=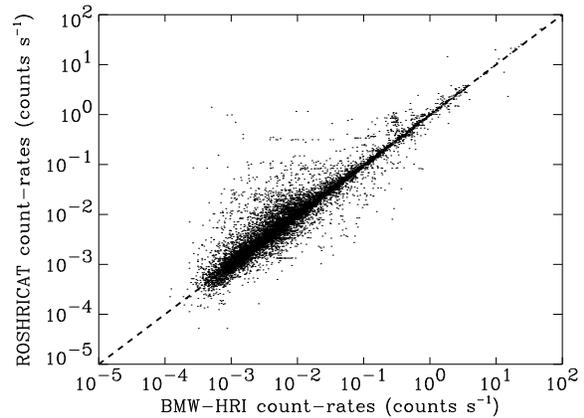,width=8cm}
\caption{ROSHRICAT - BMW-HRI: log-log plot of the 
ROSHRICAT-short (see text) count-rates versus BMW-HRI count-rates
for the 12,442 cross-correlated objects.}
\end{figure}

\subsection{Comparison with the ROSHRICAT long version}

The cross-correlation (radius of 12 arcsec) between BMW-HRI and 
ROSHRICAT-long (43,252 objects) produced 21,982 entries. 
The probability of misidentification is of the order of $3\%$ 
(that is 595 mismatches).
The number of ROSHRICAT single sources in the cross-correlation is 21,120.
Thus 22,132 
ROSHRICAT sources are not present in the BMW-HRI catalogue.
We note that the great majority of them (21,924) have S/N lower than 5.\par

There are 7,107 BMW-HRI sources without a counterpart in the ROSHRICAT-long.
Using a cross-correlation radius of 18 arcsec (i.e. corresponding to 
$3\,\sigma$ of the combined average positional error of the two catalogues) 
the number of unmatched sources reduces to 5,870. Since the number of 
spurious detections expected in the BMW-HRI is $\sim\,1,721$, we expect
that $\sim\,30\%$ of the 5,870 sources to be background fluctuations.
In order to investigate in more details these 7,107 sources, we filtered 
out sources detected in fields with extended emissions or too crowded 
on which the SASS algorithm may had problems.
Moreover, we decided not to investigate sources with $S/N <\,5$ as they are  
near the detection threshold. 
The analysis of a sub-sample of these sources with S/N $\ge\,5$ (1,170 objects)
shows that: $\sim\,49\%$ are true sources without a counterpart in the ROSHRICAT catalogue;
$\sim\,36\%$ have a ROSHRICAT association with angular separation $\ge\,18$ arcsec
while $\sim\,15\%$ are probably spurious sources (see Appendix A.3).
For clarity in Table 1 we put the results of the cross-correlations with the
short and long version of the ROSHRICAT respectively.

\begin{table*}
\begin{center}
\caption{BMW-HRI cross-correlation with ROSHRICAT catalogues.}
\begin{tabular}{|l|l|l|l|l|l|}
\hline
&&&&&\\
BMW-HRI &ROSHRICAT-short &Radius$^{b}$ &BMW-HRI            &ROSHRICAT-short     &Mismatches\\
entries &entries$^{a}$   &(arcsec)     &cross-correlations &cross-correlations  &\\
\hline
29,089  &10,708          &8            &12,442             &9,670               &73\\
\hline
\hline
BMW-HRI &ROSHRICAT-long &Radius$^{b}$ &BMW-HRI            &ROSHRICAT-long      &Mismatches\\
entries &entries$^{a}$  &(arcsec)     &cross-correlations &cross-correlations  &\\
\hline
29,089  &43,252         &12           &21,981             &21,120              &595\\
\hline
\end{tabular}
\end{center}
{(a)All the entries of the ROSHRICAT short/long version except:
sources corresponding to fields with ROR number 1 and 5; sources corresponding to 
all the fields we rejected after
or during the analysis; the 331 sources not detected by the SASS and
the 4/64 sources with a wrong declination
(see section 5.1 for more details);

(b) the radius corresponds to $2\,\sigma$ on the
source positional errors of the two catalogues.}
\end{table*}

\section{Cross-correlations with existing data-bases}

Since the sharp core of the ROSAT HRI PSF allows for a more
precise determination of the position of an X--ray source,
cross-correlations with catalogues at other wavelengths are less
affected by spurious matches.
This makes the search for counterparts much easier.  
We cross-correlated the BMW-HRI catalogue with the
largest catalogues available at other wavelengths, from radio to
optical.
For cross-correlation with catalogues at other wavelengths we used (unless otherwise
stated) a search radius of 10 arcsec.
This value comes from the major source of uncertainty in the reconstruction
of BMW-HRI source positions,which is the uncertainty in the aspect solution
of the ROSAT telescope (i.e. the boresight uncertainty).
\par

\subsection{The FIRST Survey Catalogue}

FIRST -- Faint Images of the Radio Sky at Twenty-cm -- 
covers 10,000 square degrees of the North 
Galactic Cap.
The sensitivity of the survey is of $\sim$ 1 mJy with an angular resolution of 
$\sim$ 5 arcsec (see Becker et al. \cite{becker95}). 
A catalogue containing $\sim$ 770,000 sources and covering 
$\sim$ 8,500 square degrees has been constructed 
(White et al. \cite{white97}). 
The combined sensitivity and positional accuracy of the FIRST catalogue are unprecedented 
compared with any previous wide-area radio catalogue.
FIRST source locations have an accuracy that matches or exceeds those of all currently 
available radio catalogues.
The cross-correlation with the BMW-HRI found 1,019 entries with a 
misidentification probability of the order of 2$\%$ 
(that is 18 mismatches).
Our cross-correlation is similar to the one made using FIRST and the catalogue of
X--ray sources WGACAT from ROSAT PSPC observations
(White, Giommi \& Angelini \cite{white94}; see White et al. \cite{white97}) and 
represents one of the largest
lists of X--ray/radio coincidences available to date.
In Fig. 11 we report the distributions of the angular separation ($r$ in arcsec) 
between the radio and the X--ray position for the 1,019 matched objects.

\begin{figure}
\psfig{figure=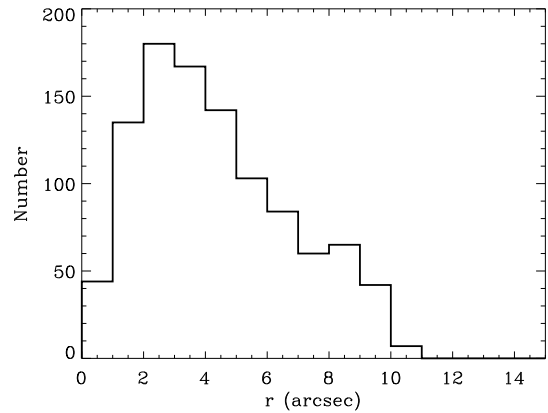,width=8cm}
\caption{FIRST - BMW-HRI: distribution of the angular separation between radio 
and X--ray positions for the 1,019 cross-correlated objects.}
\end{figure}

In Fig. 12 we plot the integrated flux densities measured in mJy 
versus X--ray flux (full column density) for the 1,019 cross-correlated sources.
The integrated flux is derived by fitting an elliptical Gaussian model to all FIRST sources. 

\begin{figure}
\psfig{figure=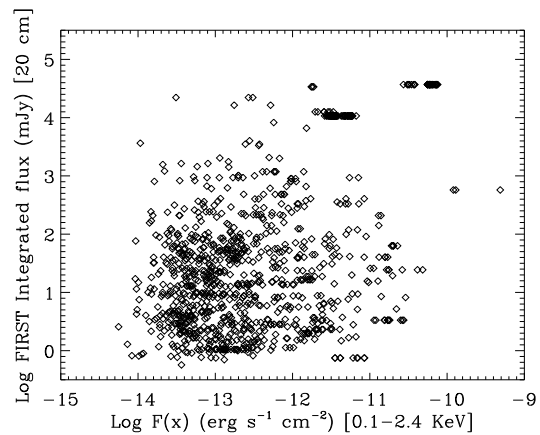,width=8cm}
\caption{FIRST - BMW-HRI: log-log plot of the integrated flux density in mJy versus X--ray flux 
(full column density) for the 1,019 cross-correlated objects.}
\end{figure}

As in the FIRST catalogue we have information about the source extension 
(the major axis, i.e. the FWHM in arcsec derived from the elliptical Gaussian 
model for the source) we plot in Fig. 13 the radio extension versus the X--ray 
one for the BMW-HRI objects classified as extended 
sources (122 entries) in the cross-correlation with FIRST. 
In the plot, X--ray extensions are the ones derived by our fitting procedure
and subtracting in quadrature the relative PSF at a given off--axis angle.
All the FIRST parameters with a brief description are reported in Appendix A
(Table A.2).

\begin{figure}
\psfig{figure=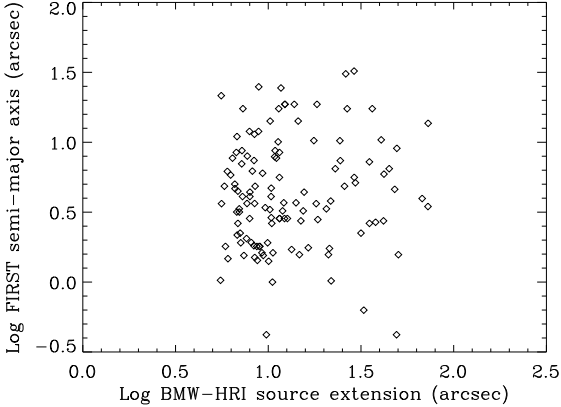,width=8cm}
\caption{FIRST - BMW-HRI: log-log plot of the major axis of the radio counterparts 
(that is the FWHM in arcsec of the elliptical Gaussian model) versus the X--ray 
extension (in arcsec) for the X--ray sources that have been classified as extended sources 
(122; see the text).}
\end{figure}
\par

\subsection{The IRAS Point Source catalogue}

The Infrared Astronomical Satellite (IRAS) conducted a survey of 98$\%$ 
of the sky, from January to November 1983, in four wavelength bands
centered at 12, 25, 60, and 100 $\mu$m leading to the
IRAS Point Source Catalogue (PSC). 
The catalogue contains some 250,000 sources (Beichman et al. \cite{beichman88}).
Away from confused regions of the sky, the PSC is complete to about 0.4, 0.5, 0.6, 
and 1.0 Jy at 12, 25, 60, and 100 $\mu$m.
The angular resolution of sources detected by IRAS varied between about 0.5 arcsec 
at 12 $\mu$m to about 2 arcmin at 100 $\mu$m.
The positional accuracy depends on source size, brightness and 
spectral energy distribution but is usually better than 20 arcsec.
Using a cross-correlation radius of 20 arcsec (because of the IRAS
positional accuracy) we found 1,149 identifications with a 
misidentification probability of $\sim\,2\%$ (20 mismatches).
We note that all the objects that have been found in common with this catalogue have been
detected in all four IRAS bands.
In Fig. 14 we plot the distributions of the angular separation ($r$ in arcsec) 
between the infrared and the X--ray position for the matched objects.

\begin{figure}
\psfig{figure=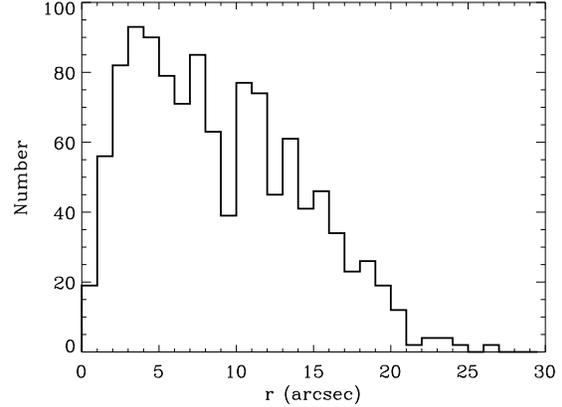,width=8cm}
\caption{IRAS - BMW-HRI: distribution of the angular separation between infrared 
and X--ray positions for the 1,149 cross-correlated objects.}
\end{figure}

Fig. 15 shows for example the 12 $\mu$m flux (in mJy) versus X--ray flux (full column 
density) for the 1,149 cross-correlated sources.
All the IRASPSC parameters with a brief description are reported in Appendix A
(Table A.3).

\begin{figure}
\psfig{figure=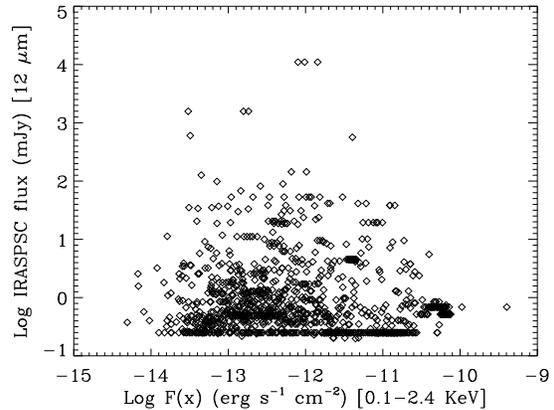,width=8cm} 
\caption{IRAS - BMW-HRI: 12 $\mu$m flux versus X--ray flux
(full column density) for the 1,149 cross-correlated objects.
}
\end{figure}

\subsection{The 2MASS Survey Catalogue}

The Two Micron All Sky Survey 
covers over 19,600 deg$^{2}$ ($\sim 50\%$) of sky 
observed from both the hemispheres.
The catalogue contains positional and photometric information for 162,213,354 
point and 585,056 extended sources observed in the three bands $J$ ($1.25\,\mu$m), 
$H$ ($1.65\,\mu$m) and $K_{s}$ ($2.16\,\mu$m). The
nominal survey completeness limits are 15.8, 15.1 and 14.3 mag respectively. 
We have cross-correlated the BMW-HRI with the 2MASS Point Source 
Catalogue 2000 (Second Incremental Release) finding 7,900 entries with a 
misidentification probability of $\sim\,28\%$ (that is 2,174 mismatches).
The number of X--ray sources found to have an infrared counterpart with
a measure in all the three bands is 7,624.

\begin{figure}
\psfig{figure=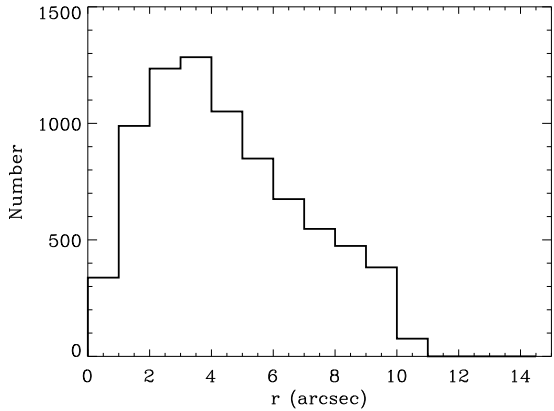,width=8cm}
\caption{2MASS - BMW-HRI: distribution of the angular separation between infrared 
and X--ray positions for the 7,900 cross-correlated objects.}
\end{figure}

In Fig. 16 we report the distributions of the angular separation ($r$ in arcsec) 
between the infrared and X--ray position for the 7,900 matched objects.
Fig. 17 shows, for example, $m_{J}$ magnitude versus X--ray 
flux (full column density) for the cross-correlated sources
with a measured $J$ magnitude (7,644).
A fraction of the cross-correlated objects, the diagonal line in Fig. 17, shows
a rather good correlation between X--ray and infrared fluxes probably due
to common dependence on the distance (see also Fig. 19 for the cross-correlation
between GSC2 and the BMW-HRI).
All the 2MASS parameters with a brief description are reported in Appendix A
(Table A.4).

\begin{figure}
\psfig{figure=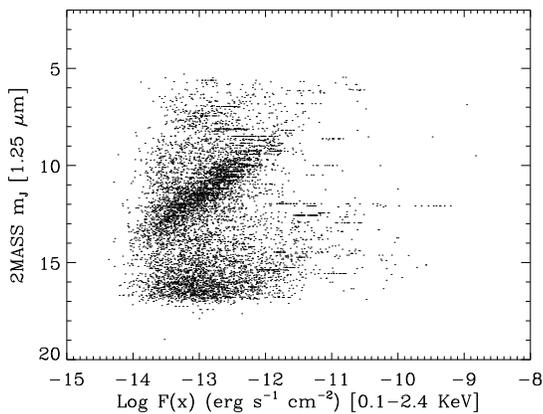,width=8cm} 
\caption{2MASS - BMW-HRI: $m_{J}$ magnitude versus logarithmic 
X--ray flux (full column density) for the cross-correlated objects.
}
\end{figure}

\begin{figure}
\psfig{figure=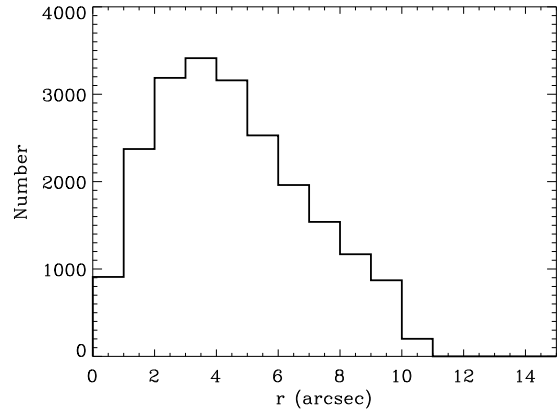,width=8cm}
\caption{GSC2 - BMW-HRI: distribution of the angular separation between 
optical and X--ray positions for the 21,311 cross-correlated objects.}
\end{figure}

\subsection{The Guide Star Catalogue 2}

As a reference for our cross-correlations in the optical, we used
the extended, yet unpublished, version of the recently released Guide Star
Catalogue  2  (GSC2).   The   GSC2  is  an  all-sky,  multi--epoch  and
multi--colour optical  catalogue based on the  digitization of $\approx$
8,000 Schmidt  plates obtained  from 13 photographic  surveys carried
out between 1953  and 1991 (McLean et al. \cite{mclean02}, in preparation).
The catalogue presently  includes more  than 1
billion  objects.  The  GSC2  is  aimed at  providing an all-sky
coverage  in  three  photographic  bands, namely  $J$, $F$  and  $N$
(roughly  comparable to Johnson $B$, $R$  and  $I$ filters). 
In  its current  version
(GSC2.2), the GSC2 covers the entire sky in
$J$ and  $F$ bands only,  while the  coverage in the  $N$ passband  is being
completed for GSC2.3.  In addition, a partial coverage in the Northern
hemisphere  and in the  South Galactic  Plane is  available in  the $V$
band (roughly comparable to Johnson $V$ filter).   
The limiting magnitudes  for GSC2 are 
$\sim\,22.5-23$, $\sim\,20-22$  and $\sim\,19.5$ in the $J$, $F$
and $N$ bands, respectively. 
In the $V$ band, the limiting magnitudes are $\sim\,19.5$ in the  
North and $\sim\,14$ in the South Galactic Plane region. 
The photometry is accurate within 0.2 magnitudes at $J\,=20$. 
The astrometry has been  calibrated 
using as a reference the  coordinates of  stars extracted from  the Tycho  
(H\o g et al. \cite{hog00}) and Hipparcos (Perryman et al. \cite{perryman97})  catalogues and 
have an absolute, intrinsic, accuracy of $\sim\,0.3$ arcsec.  
The GSC2 provides also morphological  classifications for all  the objects
detected  in at least  two bands,  with a  $\sim 90  \%$ confidence
level for  objects in at $|b|\,\ge\,5^{\circ}$ and brighter  than $J\,\sim\,19$.\par

The cross-correlation with the BMW-HRI found 21,311 entries 
indicating that an optical identification will be available for a sizeable fraction 
of BMW-HRI sources.
The mismatches are 10,508, that is a misidentification probability of $\sim\,50\%$.
This clearly indicates that the number of optical objects at the GSC2 limiting magnitude is
very high and that for many objects one finds more than one association within 10
arcsec beam-size.
In fact we found that 4,822 have two possible counterparts while 2,945 have
more than two associations.
This situation can be improved for those ROSAT HRI fields for which more than two X--ray
sources can be tied to the optical reference (i.e. boresight correction).
The total number of GSC2 non-star objects in the 21,311 cross-correlations is 9,696, while
11,331 are objects classified as ``star'' (284 are without classification).
The number of X--ray sources found to have an optical counterpart with a magnitude
measure in the J and F bands (all-sky coverage) is 13,951, while is 68 
if we also consider the N band (coverage completed for GSC2.3).

A cross-correlation with the GSC (limiting magnitude about 15) yields only 
$\sim 6,500$ objects. This indicates that the BMW-HRI source population 
has a large number of optical matching within the 15--22 magnitude range.
In Fig. 18 we report the distributions of the angular separation 
($r$ in arcsec) 
between the optical and the X--ray position for the 21,311 matched objects.

Fig. 19 shows, for example, $m_{J}$ magnitude versus X--ray flux
(full column density) for the cross-correlated sources with a measured $J$ magnitude
(18,421 objects).

\begin{figure}
\psfig{figure=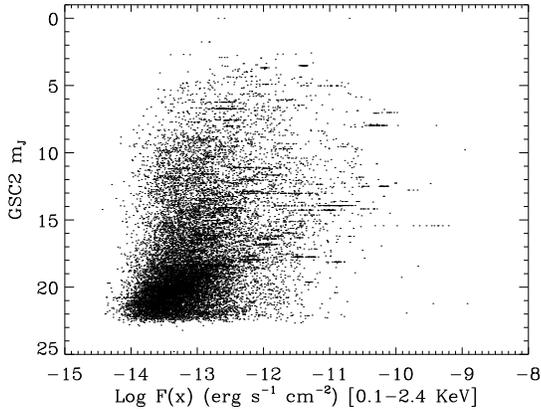,width=8cm} 
\caption{GSC2 - BMW-HRI: $m_{J}$ versus logarithmic X--ray flux (full column density) 
for the cross-correlated objects.
}
\end{figure}

\begin{figure}
\psfig{figure=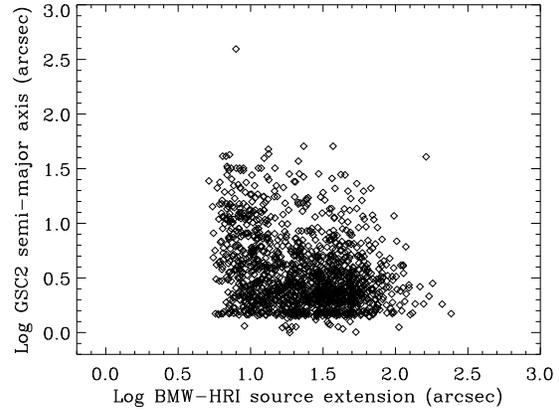,width=8cm}
\caption{GSC2 - BMW-HRI: log-log plot of the semi-major axis of the fitting ellipse (in arcsec) 
of the optical counterparts versus the X--ray extension (in arcsec) for the X--ray sources that 
have been classified as extended sources (1,470 entries; see the text).}
\end{figure}
\par

Since in the GSC2 we have an information about the source extension (the semi-major axis 
of the fitting ellipse) we plot in Fig. 20 the optical extension versus the X--ray 
extension for the BMW-HRI sources that have been classified as extended sources and  
having an optical counterpart with a measure of the semi-major axis (1,470 entries).
In the plot the X--ray extension is that derived from our fitting procedure
subtracting in quadrature the computed PSF. 
We note that 1,118 optical associations out of 1,470 have been classified as 
non-star in the GSC2, 330 as star while 22 are without a classification.
For the 330 optical objects flagged as a ``star'' in the GSC2 and with an extended
X--ray counterparts we are probably in presence of blending
sources or of low intensity sources at high off--axis in the X--ray images.
We note that 133 of them have off--axis $\ge\,10$ arcmin and that 65 of these
have S/N $<\,5$.
While the number of sources with S/N $<\,5$ is
107 out of 330.
All the GSC2 parameters with a brief description are reported in 
Appendix A (Table A.5).\par

The number of BMW-HRI sources with an association in all the
4 catalogue (FIRST, IRASPSC, 2MASS and GSC2) is 51. All but 6 are classified
as stars in the GSC2 catalogue.

\par

\begin{table*}
\begin{center}
\caption{Cross-correlation with other catalogues.}
\begin{tabular}{|l|l|l|l|l|l|l|}
\hline
&&&&&&\\
Catalogues &Wavelength &All sky fraction &Entries &Correlation &Cross-correlations &Mismatches\\
           &           &($\%$)           &        &radius (arcsec)      &                   &\\
\hline
FIRST  &20 cm  &$\sim\,25$ &$\sim\,770,000$ &10 &1,019 &18\\
\hline
IRASPSC &12-25-60-100 $\mu$m &$\sim\,98$ &$\sim\,250,000$ &20 &1,149 &20 \\
\hline
2MASS   &1,25-1.65-2.16 $\mu$m &$\sim\,50$  &162,213,354 (point) &10 &7,900 &2,174\\ 
\hline
GSC2 &J-F-N-V &all sky (J,F) &$>$ 1 billion objects &10 &21,311 &10,508\\
\hline       
\end{tabular}
\end{center}
\end{table*}

For clarity in Table 2 we report the main informations about the
FIRST, IRASPSC, 2MASS and GSC2 catalogues and the results of the cross-correlations
with the BMW-HRI.

\section{Summary}
The BMW-HRI catalogue counts all sources detected 
through a wavelet based
algorithm over the entire set of available ROSAT HRI images with exposure time greater than
100 s. 
The BMW-HRI catalogue contains 29,089 sources found in 4,303 fields with
a detection probability of $\geq\,4.2\,\sigma$.
For each source, the primary catalogue entries provide name, position, count rate, flux and
extension along with the relative errors.
The catalogue covers an area of 732 deg$^{2}$ down to a limiting flux of 
$\sim 10^{-12} \ergs \cmdue$.
The detection thresholds over the entire field of view were calculated
by means of simulations and extensive Montecarlo tests.
This allowed us to recover the log(N)-log(S) distribution for
cosmological sources down to a flux  of $\simeq 1.2 \times 10^{-14} \ergs \cmdue$
(see section 4).
The BMW-HRI catalogue has been compared with both the short and long
version of the ROSHRICAT catalogue and the cross-correlation
for the high confidence detections is at the $\sim\,90\%$ level and
at the $\sim\,71\%$ level for the long version (see appendix A).
To test the utility of the BMW-HRI catalogue in searching possible counterparts at
different wavelengths, we cross-correlated it with some of the largest existing catalogues:
the radio FIRST survey catalogue, the infrared IRAS PSC catalogue, the near-infrared 
2MASS catalogue and
the optical Guide Star Catalogue 2 (section 6).
For the radio and infrared wavelength $\sim 4\%$ of the BMW-HRI sources have an
association in the FIRST and in the IRASPSC catalogues, respectively.
While $\sim 27\%$ have a probable counterpart in the near-infrared (2MASS catalogue). 
As 2MASS survey covers $\sim\,50\%$ of sky this result is 
comparable with the one obtained in the optical (all-sky survey), from the cross-correlation 
between BMW-HRI and GSC2 ($F$ band).
If we consider all the GSC2 bands $\sim 73\%$ of the BMW-HRI sources have an optical
associations.

In order to access the service of the BMW-HRI catalogue a WEB based browser 
(via DIANA
\footnote{DIANA is a joint effort of three Italian astronomical observatories 
(Brera, Palermo and Roma) and the Italian Space Agency's Science
Data Center (ASDC). DIANA is a step towards the creation of a modern 
high energy astrophysics archive and advanced database system with 
extensive access to multiwavelength information.} 
Interface) 
with extensive access to multiwavelength information
has been built up and can be found at:

\noindent
{\tt http://www.asdc.asi.it/diana/}

The source by
coordinate environment allows the search by object name or
coordinates and to choose the output format (table only or table and
sky chart).
Full catalogue information is available 
via the HEADS - High Energy Astrophysics Database
Service on line Service (mirror of HEASARC at the Osservatorio Astronomico di Brera) with 
a remote telnet to ares.merate.mi.astro.it with ``xray'' as login (no password needed) 
and typed ``browse bmw'' to access the catalogue.\par

\begin{acknowledgements}
We thank A. Mist\`o for his help with the database software.
We are grateful to P. Giommi for discussions and M. Capalbi
for help with the WEB based browser via DIANA interface. 
This research made use of data obtained through the NASA's HEASARC at GSFC
and through the archive at MPI.
This publication makes use of data products from FIRST, IRAS, 2MASS, GSC2 surveys.
The 2MASS Survey is a joint project of the University of Massachusetts and the Infrared 
Processing and Analysis Center/California Institute of Technology, funded
by the NASA and the National Science Foundation.
The Guide Star Catalogue 2 is a joint project of the Space Telescope Science Institute
and the Osservatorio Astronomico di Torino. 
STSI is operated by the Association of Universities for Research in Astronomy, for 
the NASA under contract NAS5-26555. 
The participation of the Osservatorio
Astronomico di Torino is supported by the Italian Council for Research in Astronomy.
Additional support is provided by ESO, ST-ECF, the International GEMINI project and the ESA
Astrophysics Division.
This work was supported through Cofin, CNAA and ASI grants.
\end{acknowledgements}

\appendix

\section{Comparison between BMW-HRI and ROSHRICAT catalogues}

\subsection{Cross-correlation between BMW-HRI and ROSHRICAT-short}

In the following we report all the checks done on the 1,038 
ROSHRICAT-short sources which lack a BMW-HRI counterpart
(see sub-section 5.1).
Single cases were studied as a function of their S/N.\par

{\it (a) ROSHRICAT-short sources with S/N $\geq\,20$ (20 objects)}. 
We found that all but one of the sources have a counterpart in the BMW-HRI catalogue.
If we extend the cross-correlation radius we found that 18 sources have a counterpart 
within $\sim\,20$ arcsec and one within $\sim\,30$ arcsec.
We checked each of these source to investigate possible mismatches.
For the remaining source we checked that it is a 
spurious detection near a very bright source (with a positional shift of 1.5 arcmin).\par

{\it (b) ROSHRICAT-short sources with S/N between 5 and 20 (358 objects)}.
To minimize the number of mismatches, we
cross-correlated these sources using a radius of 18 arcsec,
corresponding to the $3\,\sigma$ on the combined positional error of the two
catalogues. 
The cross-correlation found 369 BMW-HRI sources corresponding to 221
ROSHRICAT sources, i.e., 137 ROSHRICAT objects have no counterpart
in our catalogue.
We checked 20$\%$ of these sources (28 objects) finding that:
(i) 47$\%$ are spurious detections close to bright
sources; 
(ii) 32$\%$ are associated with extended emission and likely spurious 
since the SASS detection algorithm is not suited to the detection
of extended sources; 
(iii) 7$\%$ are from a blending of two nearby point-like sources; 
(iv) 14$\%$ are detected with a S/N lower than what reported by the SASS
and we recover them if we use a lower detection threshold 
(detection probability of $3.5\,\sigma$ instead of $4.2\,\sigma$).\par

{\it (c) ROSHRICAT-short sources with S/N $<$ 5 (660 objects)}.
These sources are rather faint and 16 have S/N lower than 4 even in the ROSHRICAT-short
itself. 
Their detection depends critically on the extraction radius and on local
background.
The great majority should have been missed by our detection algorithm
because they are indeed too faint to satisfy our detection threshold
(which is not based on S/N ratio).\par

\subsection{Check on the 331 obvious sources not detected by the SASS and manually added}

We found that 79 of these sources are in fields we discarded (ROR number 1 and 5 
or fields rejected during the analysis).
We found that 231 of the 252 remaining sources  
have a counterpart in our catalogue: 230 within 30 arcsec and 1 within 50 arcsec.
We checked with a visual inspection the remaining 21 obvious sources finding that:
8 are spurious detections; 8 are very faint sources which we are able to detect if we use 
a lower detection threshold for our algorithm (significance $\geq\,3.5\sigma$); 3 sources are
ambiguous since XIMAGE detects only a 
$3\,\sigma$ upper limit; one source is at the edge and can be detected only by
integrating its flux over all the HRI energy channels (see section 2); the
remaining one is very close to an hot spot and was classified as
spurious.
For the 8 spurious we note that 5 of them are not far away from bright sources.
It could be that the manual insertion of the coordinates were wrong.

\subsection{Cross-correlation between BMW-HRI and ROSHRICAT-long}

There are 7,107 BMW-HRI sources without a counterpart in the ROSHRICAT-long (see sub-section 5.2). 
Using a cross-correlation radius of 18 arcsec
(i.e. corresponding to $3\,\sigma$ on the combined average positional error of
the  two catalogues) the number of unmatched sources reduces to
5,870. 
For the visual inspection we excluded sources in fields with large
extended emission and too crowded and we
selected a sample consisting of 1,170 sources with S/N $\geq\,5$.
These sources were considered and studied as a function of their S/N.\par

{\it (a) BMW-HRI sources with S/N $\geq\,20$ (27 objects)}:
(i) 63$\%$ are bright or
relatively bright objects (13 at the detector edge and 4 are targets);
(ii) 26$\%$ all with off--axis $>\,17$ arcmin 
have a ROSHRICAT counterpart with an angular separation 
larger than 18 arcsec;
(iii) the remaining 11$\%$ are at the edge of the detector (off--axis $\geq\,18$ arcmin) 
and are hardly recognizable at a visual inspection.\par

{\it (b) BMW-HRI sources with S/N between 5 and 20 (1,143 objects)}:
we checked 10$\%$ of these sources. 
We note that 86$\%$ of these have an off--axis angle
$\geq\,14$ arcmin and 40$\%$ are extended sources.
We found that: (i) 36$\%$ have a counterpart
in the ROSHRICAT-long with an angular separation $>\,18$ arcsec.
The combination of extended emission with high off--axis
angle can explain the highest angular separation.
In fact almost all sources with angular separation $>\,30$
arcsec have off-axis $\geq\,14$ arcmin and are extended sources;
(ii) 48$\%$ are sources without a counterpart in the
ROSHRICAT-long and 50$\%$ of these are
well visible at a visual inspection.
We note that 85$\%$ have off--axis $\geq\,14$ arcmin
and that 42$\%$ are extended sources;
(iii) 16$\%$ are
sources of ambiguous interpretation: all these sources but one have
off--axis $\geq\,17$ arcmin and seem fluctuations of the background 
at a visual inspection.
We note that 11$\%$ of these are extended sources.
If we use the XIMAGE package at the positions of the point sources
(16 objects) we find a $3\,\sigma$ upper limits for 11 sources
while for the remaining 5 the algorithm detects sources
with very low signal to noise ratio: 4 out 5 with S/N
$\leq\,2$ and the remaining with S/N $\sim\,6$.

\vfill

\begin{table*}
\caption{BMW-HRI parameters.}
\begin{center}
\begin{tabular}{|l|l|}
\hline
&\\
BMW Parameter &Description\\
\hline
Source Name  &name of cataloged detection following IAU conventions,  
e.g. 1BMW143615.8+524825\\
\hline
RA &source right ascension (J2000, hhmmss.)\\
\hline
DEC &source declination (J2000, ddmmss.)\\
\hline
RA error &error in RA (arcsec)\\
\hline
DEC error &error in DEC (arcsec)\\
\hline
Tot error &total positional error (arcsec)\\
\hline
JRA &source RA (J2000, degrees)\\
\hline
JDEC &source RA (J2000, degrees)\\
\hline
LII &source Galactic Longitude (degrees)\\
\hline
BII &source Galactic Latitude (degrees)\\
\hline
Rebin &rebin of the subimage in which the source was found\\
\hline
X pixel &source x coordinate (pixels)\\
\hline
Y pixel &source y coordinate (pixels)\\
\hline
Offaxis &offset from Target position (arcmin)\\
\hline
Field &field name\\
\hline
Target RA &target RA (J2000, hhmmss.)\\
\hline
Target DEC &target DEC (J2000, ddmmss.)\\
\hline
ROR &ROR number (e.g. 200005)\\
\hline
Sequence &observation sequence id (e.g. rh200005a00)\\
\hline
Start obs date &start observation date (ddmmyy)\\
\hline
Start obs time &start observation time (hh:mm:ss.) \\
\hline
End obs date &end observation date\\
\hline
End obs time &end observation time\\
\hline
Exposure &exposure live time (seconds)\\
\hline
Map exposure &exposure from the exposure map (seconds)\\
\hline
Num reb1 &number of sources detected at reb=1\\
\hline
Num reb3 &number of sources detected at reb=3\\
\hline
Num reb6 &number of sources detected at reb=6\\
\hline
Num reb10 &number of sources detected at reb=10\\
\hline
SNR &signal-to-noise ratio for detection\\
\hline
Wavelet SNR &signal-to-noise ratio in the wavelet space\\
\hline
Probability &probability for detection\\
\hline
Count rate &estimate of net count rate (CR, count s$^{-1}$)\\
\hline
Fit CR error &count rate error from the fit procedure (count s$^{-1}$)\\
\hline
Stat CR error &count rate error from statistics (count s$^{-1}$)\\
\hline
CR error &count rate error (maximum between fit CR error and stat CR error)\\
\hline
Total counts &total counts detected (CR by exposure)\\
\hline
Vignetting &vignetting correction\\
\hline
PSF &PSF correction\\
\hline
Counted CR &source CR from the counting procedure\\
\hline
Counted CR error &source counted CR error\\
\hline
Background &background count rate\\
\hline
Near sources &number of sources within 2 sigma from the detected source\\
\hline
Nh &weighted average nH (cm$^{-2}$)\\
\hline
Conversion Factor1 (CF1) &conversion factor from null column density 
(erg cm$^{-2}$ count$^{-1}$)\\
\hline
Fx1 &HRI X--ray flux from CF1\\
\hline
Conversion Factor1 (CF2) &conversion factor from full column density
(erg cm$^{-2}$ count$^{-1}$)\\
\hline
Fx2 &HRI X--ray flux from CF2\\
\hline
Extension &sigma source in original pixels (arcsec)\\
\hline
Extension error &sigma source error\\
\hline
Flag sigma &fixed sigma? (Y/N)\\
\hline
Flag extended &extended sources? (POINT/EXTENDED)\\
\hline
Extension significativity &significativity of the extension\\
\hline
Fit $\chi^{2}$ &$\chi^{2}$ from the fit procedure\\
\hline
Cat version &catalogue version\\
\hline
SASS version  &version of the SASS processing system\\
\hline
\end{tabular}
\end{center}
\end{table*}

\begin{table*}
\caption{FIRST parameters.}
\begin{center}
\begin{tabular}{|l|l|}
\hline
&\\
FIRST Parameter &Description\\
\hline
FIRST name &FIRST source name (e.g. 11030$+$38071E)\\
\hline
RA &FIRST source RA (J2000, degrees)\\
\hline
DEC &FIRST source DEC (J2000, degrees)\\
\hline
W &warning flag for sidelobe source\\
\hline
Pflux &peak flux density (mJy)\\
\hline
Iflux &integrated flux density (mJy)\\
\hline
Rms &local noise at the source position (mJy)\\
\hline
Maj &major axis after deconvolution (FWHM in arcsec; elliptical Gaussian model)\\
\hline
Min &minor axis after deconvolution (in arcsec)\\
\hline
PA &position angle after deconvolution (in arcsec; degrees east of north)\\
\hline
fMaj &measured major axis (arcsec)\\
\hline
fMin  &measured minor axis (arcsec)\\
\hline
fPA &measured position angle (arcsec)\\
\hline
FIRSTBMW &angular distance between FIRST and X--ray position (arcsec)\\
\hline
FIRST comp &number of FIRST cross-correlations\\
\hline
BMWFIRST version &BMW-FIRST cross-correlation version\\
\hline
\end{tabular}
\end{center}
\end{table*}

\begin{table*}
\caption{IRASPSC parameters.}
\begin{center}
\begin{tabular}{|l|l|}
\hline
&\\
IRASPSC Parameter &Description\\
\hline
IRASPSC name &source IRASPSC name (e.g. 18352+3844)\\
\hline
RA &IRASPSC RA source (J2000, degrees)\\
\hline
DEC &IRASPSC DEC source (J2000, degrees)\\
\hline
F12 &12 micron flux (mJy)\\
\hline
F25 &25 micron flux (mJy)\\
\hline
F60 &60 micron flux (mJy)\\
\hline
F100 &100 micron flux (mJy)\\
\hline
IRASPSCBMW &angular distance between IRAS and X--ray position (arcsec)\\
\hline
IRASPSC comp &number of IRASPSC cross-correlations\\
\hline
IRASPSC version &BMW-IRASPSC cross-correlation version\\
\hline
\end{tabular}
\end{center}
\end{table*}

\begin{table*}
\caption{2MASS parameters.}
\begin{center}
\begin{tabular}{|l|l|}
\hline
&\\
2MASS Parameter &Description\\
\hline
RA &2MASS RA source (J2000, degrees)\\
\hline
DEC &2MASS DEC source (J2000, degrees)\\
\hline
Jmag &2MASS J magnitude\\
\hline
Jmag error &2MASS J magnitude error\\
\hline
Hmag &2MASS H magnitude\\
\hline
Hmag error &2MASS H magnitude error\\
\hline
Kmag &2MASS K magnitude\\
\hline
Kmag error &2MASS K magnitude error\\
\hline
2MASSBMW &angular distance between 2MASS and X--ray position (arcsec)\\
\hline
2MASS comp &number of 2MASS cross-correlations\\
\hline
BMW2MASS version &BMW-2MASS cross-correlation version\\
\hline
\end{tabular}
\end{center}
\end{table*}

\vfill
\hfill

\begin{table*}
\caption{GSC2 parameters.}
\begin{center}
\begin{tabular}{|l|l|}
\hline
&\\
GSC2 Parameter &Description\\
\hline
GSC2 name &GSC2 source name (e.g. N2320203180)\\
\hline
RA &GSC2 source RA (J2000, degrees)\\
\hline
DEC &GSC2 source DEC (J2000, degrees)\\
\hline
RA error &source RA error (degrees)\\
\hline
DEC error &source DEC error (degrees)\\
\hline
GSC2 epoch &GSC2 epoch (e.g. 1991.329956)\\
\hline
raPM &RA proper motion (mas yr$^{-1}$)\\
\hline
decPM &DEC proper motion (mas yr$^{-1}$)\\
\hline
raPMerr &RA proper motion error (mas yr$^{-1}$)\\
\hline
decPMerr &DEC proper motion error (mas yr$^{-1}$)\\
\hline
Fmag &F magnitude\\
\hline
Fmag error &F magnitude error\\
\hline
Jmag &J magnitude\\
\hline
Jmag error &J magnitude error\\
\hline
Vmag &V magnitude\\
\hline
Vmag error &V magnitude error\\
\hline
Nmag &N magnitude\\
\hline
Nmag error &N magnitude error\\
\hline
a &semi-major axis\\
\hline
e &eccentricity\\
\hline
PA &position angle (arcsec)\\
\hline
c &class code (0=star,1=galaxy,2=blend,3=non-star,4=unclassified,5=defect)\\
\hline
GSC2BMW &angular distance between GSC2 and X--ray position (arcsec)\\
\hline
GSC2 comp &number of GSC2 cross-correlations\\
\hline
BMWGSC2 version &BMW-GSC2 cross-correlation version\\
\hline
\end{tabular}
\end{center}
\end{table*}

\end{document}